\documentclass[8.5pt,twoside,twocolumn]{article}
\usepackage[super,sort&compress,comma]{natbib} 
\usepackage[version=3]{mhchem}
\usepackage{times,mathptmx}
\usepackage{sectsty}
\usepackage{balance} 

\usepackage{graphicx} 
\usepackage{lastpage}
\usepackage{amsmath}
\usepackage{amssymb}
\newcommand{\un}[1]{\ensuremath{\unskip\,\mathrm{#1}}}
\newcommand{\form}{\ce{C12EO5}}
\newcommand{\vect}[1]{\mathbf{#1}}
\newcommand{\lpp}{$\lambda_\text{max}$}
\usepackage[format=plain,justification=raggedright,singlelinecheck=false,font=small,labelfont=bf,labelsep=space]{caption} 
\usepackage{fancyhdr}
\begin{document}

\renewcommand{\thefootnote}{\fnsymbol{footnote}}
\setcounter{secnumdepth}{5}

\makeatletter 
\def\subsubsection{\@startsection{subsubsection}{3}{10pt}{-1.25ex plus -1ex minus -.1ex}{0ex plus 0ex}{\normalsize\bf}} 
\def\paragraph{\@startsection{paragraph}{4}{10pt}{-1.25ex plus -1ex minus -.1ex}{0ex plus 0ex}{\normalsize\textit}} 
\renewcommand\@biblabel[1]{#1}            
\renewcommand\@makefntext[1]%
{\noindent\makebox[0pt][r]{\@thefnmark\,}#1}
\makeatother 
\renewcommand{\figurename}{\small{Fig.}~}
\sectionfont{\large}
\subsectionfont{\normalsize} 

\fancyfoot{}
\fancyfoot[CO]{\thepage} 
\fancyfoot[CE]{\thepage} 
\renewcommand{\headrulewidth}{1pt} 
\renewcommand{\footrulewidth}{1pt}
\setlength{\arrayrulewidth}{1pt}
\setlength{\columnsep}{6.5mm}
\setlength\bibsep{1pt}

\twocolumn[
  \begin{@twocolumnfalse}
\noindent\LARGE{\textbf{Infrared dichroism of gold nanorods controlled using a magnetically addressable mesophase}}
\vspace{0.6cm}

\noindent\large{\textbf{Kostyantyn Slyusarenko,\textit{$^{a}$} Doru Constantin,$^{\ast}$\textit{$^{a}$} Benjamin Ab\'{e}cassis,\textit{$^{a}$} Patrick Davidson,\textit{$^{a}$} and Corinne Chan\'{e}ac\textit{$^{b}$}}}\vspace{0.5cm}

\noindent{Published in \textit{J. Mater. Chem. C} \textbf{2}(26), 5087-5092 (2014).}

\noindent{\textbf{DOI: 10.1039/C4TC00318G}}
\vspace{0.6cm}

\noindent \normalsize{Gold nanorods have unique optical properties, which make them promising candidates for building nano-structured materials using a ``bottom-up'' strategy. We formulate stable bulk materials with anisotropic optical properties by inserting gold and iron oxide nanorods within a lamellar mesophase. Quantitative measurements of the order parameter by modelling the absorbance spectra show that the medium is macroscopically aligned in a direction defined by an external magnetic field. Under field, the system exhibits significant absorption dichroism in the infrared range, at the position of the longitudinal plasmon peak of the gold nanorods (about 1200~nm), indicating strong confinement of these particles within the water layers of the lamellar phase. This approach can yield soft and addressable optical elements.}
\vspace{0.5cm}
 \end{@twocolumnfalse}
  ]

\footnotetext{\textit{$^{a}$~Laboratoire de Physique des Solides, Univ. Paris-Sud,
CNRS, UMR8502, 91405 Orsay Cedex, France Fax: 01 6915 6086; Tel: 01 6915 5394; E-mail: doru.constantin@u-psud.fr}}
\footnotetext{\textit{$^{b}$~Laboratoire de Chimie de la Mati\`{e}re Condens\'{e}e,
Univ. Paris VI, CNRS, UMR7574, 75252 Paris Cedex 05, France}}

\section{Introduction}

Orienting anisotropic nanoparticles at the bulk level is a current challenge in nanotechnology, being a necessary step to exploit the anisotropic physical properties of individual nanoparticle into devices. This is particularly relevant for noble metal particles, actively used due to their particular optical properties. The absorption spectrum of such particles is dominated by peaks corresponding to the surface plasmon modes. For gold in water, a first peak occurs at a wavelength of $\lambda \simeq 520 \un{nm}$ for spheres and for the transverse mode of nanorods, while the longitudinal plasmon peak (LPP) of the latter shifts as a function of particle size and geometry \cite{Novotny07}. At high aspect ratio (AR), the LPP wavelength {\lpp} increases nearly linearly with the particle length.

Among the host materials used for nanoparticle self-assembly, surfactant mesophases doped with inorganic nanoparticles \cite{Constantin:2014} are easily produced in bulk, and thus especially adapted to the formulation of nano-structured materials via bottom-up techniques. Moreover, their orientational order allows us to take advantage of the anisotropic properties of asymmetric particles, by propagating this anisotropy up to the macroscopic level of the bulk material (\textit{e.g.} optical birefringence and dichroism.) In recent years, gold nanoparticles (both spheres and rods) were inserted within nematic or lamellar mesophases and their optical response was thoroughly studied. 

However, bulk self-assembled systems based on gold nanorods so far had their LPP in the visible \cite{Liu10,Liu12} or near-infrared range (below about 900~nm) \cite{Sreeprasad08,Xiao12}; to the best of our knowledge, there is no study yet reporting the macroscopic orientation of gold nanorods having their LPP in the infrared domain. In the present study we focus on wavelengths around 1200~nm. This spectral range is particularly interesting for applications in two different domains:
\begin{itemize}
\item{} Human skin and subcutaneous tissue exhibit here the highest penetration depth within the visible and infrared spectrum\cite{Bashkatov05}, facilitating both imaging and photothermal treatment using metal nanoparticles\cite{Huang06}.
\item{} Several popular types of laser (Nd:YAG, HeNe and Yb-doped fibers \cite{Jeong04}) operate in this range. It is therefore essential to develop suitable optical elements (such as polarizers).
\end{itemize}

We synthesized, functionnalized and guided the self-organization of the nanorods by a soft matter template (a lamellar mesophase of nonionic surfactant). In this type of phase, layers of water separated by surfactant bilayers can be aligned over millimeter ranges. Furthermore, the water layer thickness can be finely tuned by the composition of the phase, between about 1 and 100~nm. We demonstrated by infrared dichroism that the gold nanorods are strongly confined within the lamellar matrix. Adding iron oxide nanorods with magnetic properties renders the phase sensitive to an applied magnetic field, so that its orientation (and thus its optical anisotropy) can be controlled at will.

\section{Experimental}

\begin{figure*}[htbp]
\includegraphics[width=0.5\textwidth,angle=0]{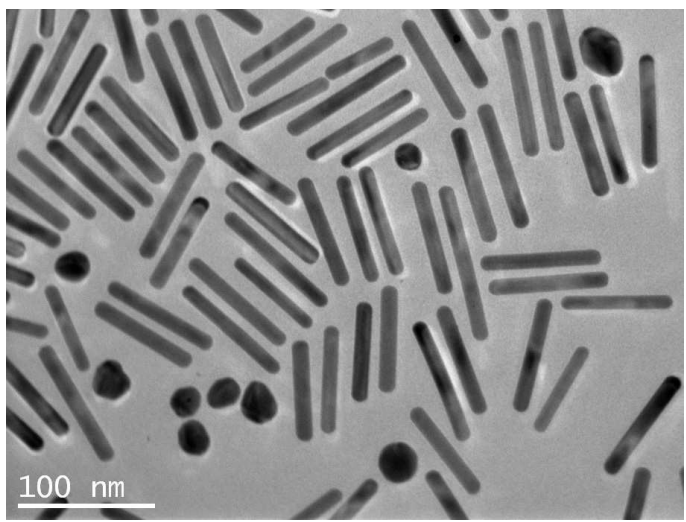}%
\includegraphics[width=0.5\textwidth,angle=0]{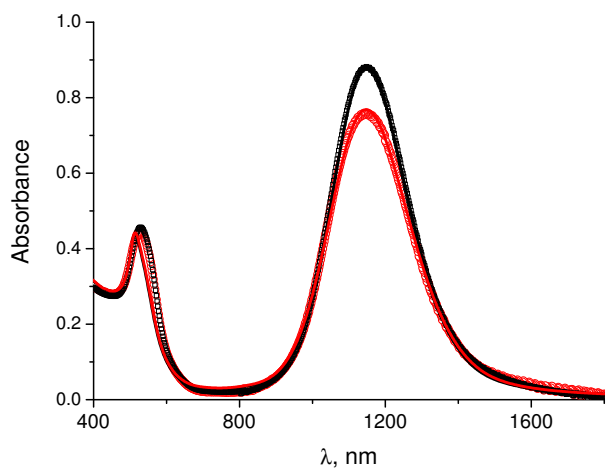}
\begin{center}
a) \hspace*{10cm} b)
\end{center}
\caption{a) Transmission electron microscopy (TEM) image of the gold nanorods. b) Absorbance of the gold nanorods dispersed in water (red dots) and confined in the lamellar matrix (black squares). The symbols are the experimental values and the curves are fits with Equation~\eqref{Eq:Abs}. We fix $S=0$ for the aqueous solution.}
\label{fig:Abs}
\end{figure*}

\subsection{Gold nanorod synthesis}

All reagents were from Sigma Aldrich, used as received.

A stock suspension of gold nanoparticles was synthesized by a modified seed-growth method \cite{Ye:2013}. The seed solution was prepared by mixing \ce{NaAuCl4} solution (5~ml, 0.5~mM) with  hexadecyltrimethylammonium bromide (CTAB) solution (5~ml, 0.2~M). 0.6~ml of fresh 0.01 M~\ce{NaBH4} solution were injected into the Au-CTAB solution under stirring at 1200~rpm for 2~min at 30$ ^{\circ}$C, which resulted in the formation of a brownish-yellow solution. The solution was kept undisturbed at  30$ ^{\circ} $C for 30~min before use.

The growth solution was prepared by dissolving 9~g of CTAB with 1.23~g of sodium oleate in 250~ml of water at 30~$^{\circ}$C. Then 30~ml of 4~mM of \ce{AgNO3} solution were added. The mixture was kept undisturbed for 15~min, after which 250~ml of 1~mM \ce{HAuCl4} solution were added and when the solution became colorless 62~ml of HCl (37 w.\% in water) were injected to lower the pH. After 15 min of slowly stirring at 400 rpm, 1.25~ml of 64~mM ascorbic acid were added to the solution which was vigorously stirred for 30~sec. 

In the final step, 0.8~ml of seed solution was added to the growth solution. The resulting solution was stirred at 600~rpm for 30~s and left at 30~$^\circ$C for 12~h. The color of the solution changes with time depending on the aspect ratio of the gold nanorods. The suspension also contains a significant amount of spheres and cubes (volume fraction comparable to that of the nanorods), but their contribution to the infrared absorbance is negligible.

Pure water was added to the final solution to stop the synthesis and to reach a CTAB concentration of 50~mM. The stock suspension of gold nanoparticles was concentrated by centrifuging at 6000~g for 30~min and removing the supernatant. 

Due to the presence of CTAB in the initial suspension of gold nanorods, the latter aggregate upon mixing with the lamellar phase. We stabilized the system by exchanging the CTAB with the nonionic surfactant (1-mercaptoundec-11-yl)hexa(ethylene glycol) (MUDOL). First, the gold suspension was diluted 100 times, to a CTAB concentration of 0.5~mM and $\phi_\text{Au} = 0.01 \un{vol \%}$. Then, 10~$\mu$l of 5~mM MUDOL were added to 0.5~ml of the dilute suspension. The mixture was vigorously stirred for 1~min and left for 24 h at 30~${^\circ}$C. This waiting time is sufficient to complete the CTAB to MUDOL exchange at the surface of the nanorods \cite{Hamon:2012}.

\subsection{Goethite nanorod synthesis}
The goethite ($\alpha-\un{FeOOH}$) nanorods were synthesized according to
well-established protocols \cite{Atkinson67,Jolivet04}. They are lath-shaped, with length $L=350 \pm 120 \, \un{nm}$ and width $W=36 \pm 13  \, \un{nm}$, as determined by TEM. Stable aqueous suspensions of goethite nanorods are obtained by repeated centrifugation and dispersion in water at $\un{pH} =3$, where their surface is hydroxylated, with a surface charge of $0.2 \un{C \, m^{-2}}$ (the isoelectric point corresponds to $\un{pH} = 9$). Although bulk goethite is antiferromagnetic, the nanorods bear along their long axis a permanent magnetic dipole $\mu$ of a few thousand $\mu_{B}$ (with $\mu _{B} = 9.274 \times 10^{-24}\, \un{J/T}$ the Bohr magneton), probably due to uncompensated surface spins. Therefore, in suspension, the nanorods are easily aligned parallel to a small magnetic field. Furthermore, the easy magnetisation axis is perpendicular to this direction so that, at high applied fields, the induced magnetic moment overtakes the permanent one and the orientation of the rods switches to perpendicular to the field at a critical value $H \sim 350 \un{mT}$ \cite{Lemaire02}.

\subsection{Lamellar phase}
The matrix is the \form/hexanol/\ce{H2O} system, with {\form} the nonionic surfactant penta(ethylene glycol) monododecyl ether. Its lamellar phase can be diluted down to spacings $d$ in the micron range, while the bilayer thickness $\delta \approx 2.9 \un{nm}$
\cite{Freyssingeas96,Freyssingeas97}. We used a hexanol/{\form} ratio of 0.35 by weight, corresponding to a molar ratio of 1.3 (hexanol
molecules for each surfactant molecule). The main role of hexanol is to bring the lamellar phase domain down to room temperature. The
surfactant was acquired from Nikko and the hexanol from Fluka; they were used without further purification.

\subsection{Magnetic field alignment}
The magnetic field was applied using a home-made setup based on permanent magnets with a variable gap. One can thus reach field intensities of up to 0.9~T. During field application the samples were observed using an Olympus BX51 microscope (with $5\times$--$40\times$ objectives) using linearly polarized light and, when specified, an analyzer perpendicular to the incident polarization.

\subsection{Absorbance spectroscopy} \label{subsec:absorb}
We used a Cary~5000 spectrometer (Agilent). The beam was linearly polarized by a rotating UV polarizer (Edmund Optics) and defined by a $1 \! \times 4 \, \un{mm^2}$ slit placed immediately before the sample.

The individual absorbance of a nanorod oriented along $\vect{u}$ illuminated by a light beam polarized along $\vect{p}$ can be written as:
\begin{equation}
a(\lambda,\vect{u})=a_\text{long}(\lambda) (\vect{p} \cdot \vect{u})^2 + a_\text{tr}(\lambda) [1-(\vect{p} \cdot \vect{u})^2]
\label{Eq:orient}
\end{equation}

If the nematic director $\vect{n}$ of the nanorod population is parallel to the direction of the beam (and hence perpendicular to $\vect{p}$) as in Figure~\ref{fig:POM}(a), we can write $\left \langle (\vect{p} \cdot \vect{u})^2 \right \rangle = \left \langle \frac{1-(\vect{u} \cdot \vect{n})^2}{2} \right \rangle $. The macroscopic absorbance, averaged over $f(\vect{u})$, is then:
\begin{equation}
\label{Eq:Aavg}
\begin{split}
&A(\lambda) \propto \left \langle a(\lambda,\vect{u}) \right \rangle  =\\ &a_\text{long}(\lambda) \left \langle \frac{1-(\vect{u} \cdot \vect{n})^2}{2} \right \rangle +
a_\text{tr}(\lambda) \left \langle \frac{1+(\vect{u} \cdot \vect{n})^2}{2} \right \rangle \Rightarrow \\
&A(\lambda) \propto \dfrac{1-S}{3} a_\text{long}(\lambda) + \dfrac{2+S}{3} a_\text{tr}(\lambda)
\end{split}
\end{equation}
where in the last step we used \eqref{Eq:Order_Def} in order to finally obtain \eqref{Eq:Abs}. By symmetry, the relation \eqref{Eq:Aavg} holds for all polarizations $\vect{p}$ with $\vect{p} \bot \vect{n}$ and hence for any polarization state of the beam in Figure~\ref{fig:POM}(a).

If, on the other hand, $\vect{n}$ is perpendicular to the beam direction (Figure~\ref{fig:POM}(b)) one must consider the two absorbance functions, $A_{\|}(\lambda)$ and $A_{\bot}(\lambda)$, for $\vect{p} \| \vect{n}$ and $\vect{p} \bot \vect{n}$, respectively. In the first case (corresponding to $\beta = 90^{\circ}$ in Figure~\ref{fig:POM}(b)) we simply have $(\vect{p}_{\|} \cdot \vect{u})^2 = (\vect{u} \cdot \vect{n})^2$ and
\begin{equation}
\label{Eq:Apara}
A_{\|}(\lambda) \propto \dfrac{1+2S}{3} a_\text{long}(\lambda) + \dfrac{2-2S}{3} a_\text{tr}(\lambda)\, ,
\end{equation}
while in the second one (for $\beta = 0^{\circ}$) $\left \langle (\vect{p}_{\bot} \cdot \vect{u})^2 \right \rangle = \left \langle \frac{1-(\vect{u} \cdot \vect{n})^2}{2} \right \rangle$, yielding:
\begin{equation}
\label{Eq:Aperp}
A_{\bot}(\lambda) \propto \dfrac{1-S}{3} a_\text{long}(\lambda) + \dfrac{2+S}{3} a_\text{tr}(\lambda)\, .
\end{equation}
The simplifying assumption $a_\text{tr}(\lambda_{\text{max}}) \ll a_\text{long}(\lambda_{\text{max}})$ then leads to \eqref{Eq:Sexp}.

\section{Results and discussion}

\subsection{Sample preparation}

To reach the infrared domain we used nanorods with an aspect ratio $AR = 7.6 \pm 0.6$, and $\lambda_\text{max} \approx 1180 \un{nm}$, see Figure~\ref{fig:Abs}. They were obtained using the classical seed synthesis recently modified by Murray's group \cite{Ye:2013}. For such large and anisotropic particles, the challenge is obtaining both a stable dispersion and a significant confinement so that the particles can achieve a high degree of orientational order. To reach these goals we functionalize the gold nanorods with a nonionic surfactant very similar in structure to the mesophase surfactant.

Afterwards, we prepared lamellar phases of {\form}, hexanol and gold nanorods by mixing appropriate amounts of stock particle solutions with a concentrated lamellar phase and by adding water as necessary (this particular mixing sequence affords good control of the hexanol/{\form} ratio, which strongly influences the phase diagram of the system). The typical volume fractions are: $\phi_m = (V_{\form} + V_{\text{hexanol}})/V_{\text{total}} = 6.3 \un{vol \%}$ for the membrane, corresponding to a smectic repeat distance $d \simeq 42 \text{nm}$, and $\phi_\text{Au} = 0.01 \un{vol \%}$ for the gold nanorods.

\begin{figure}[htbp]
\includegraphics[width=0.5\textwidth,angle=0]{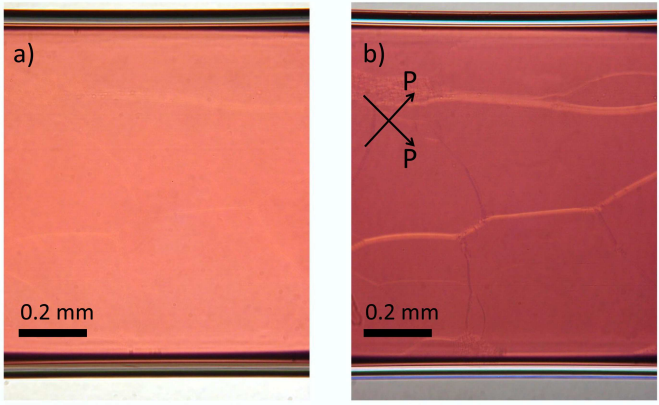}
\caption{Microscopy images of the lamellar phase doped with gold nanorods in natural light (a) and between crossed polarizers (b).  The double arrows labelled ``P'' represent the directions of the polarizer and analyzer. The exposure time for image (b) was 250 times longer than for (a).}
\label{fig:Image}
\end{figure}
The samples were introduced in flat glass capillaries, 50~$\mu$m thick and 1~mm wide (Vitrocom, NJ, USA) by gentle suction with a hand-operated 1~ml syringe. The aspect of the samples immediately after preparation is shown in Figure~\ref{fig:Image}, in natural light and between crossed polarizers. The uniform red color of the capillary shows that the distribution of the nanorods is homogeneous. Furthermore, the sample has remained stable for months. The as-prepared doped lamellar phase is not birefringent, as illustrated by the very low intensity transmitted between crossed polarizers (see Figure~\ref{fig:Image}b). This indicates that the lamellar phase orients homeotropically, \textit{i.e.} with the director $\vect{n}$ perpendicular to the flat faces of the capillary.

In order to determine the degree of orientation of the gold nanorods in a more quantitative manner, we derived a theoretical expression for the absorbance and used it to adjust the experimental spectra. The gold nanorods are symmetrically distributed with respect to $\vect{n}$, forming a uniaxial nematic distribution $f(\vect{u)}$ (where $\vect{u}$ represents the nanorod orientation) with director along $\vect{n}$ and with an order parameter $S$ given by
\begin{equation}
\label{Eq:Order_Def}
S =  \left \langle \frac{3(\vect{u} \cdot \vect{n})^2 - 1}{2} \right \rangle  = \int \text{d}\vect{u} f(\vect{u}) \frac{3(\vect{u} \cdot \vect{n})^2 - 1}{2} \, .	
\end{equation}
Particles perfectly aligned along one axis (the director $\vect{n}$) have an order parameter $S=1$, while $S=-0.5$ if they are perpendicular to this axis. If the orientation is completely isotropic (constant $f(\vect{u})$) $S=0$. The absorbance of an electromagnetic beam propagating along $\vect{n}$ by a suspension of gold nanorods with order parameter $S$ is (see \S~\ref{subsec:absorb}):
\begin{equation}
A(\lambda) \propto (1-S)a_\text{long}(\lambda) + (2+S)a_\text{tr}(\lambda) \, ,
\label{Eq:Abs}
\end{equation}
where $a_\text{tr}$ and $a_\text{long} $ are the absorbances of a nanorod due to the transverse and longitudinal plasmonic resonances (for an electric field perpendicular and parallel to the nanorod, respectively), calculated according to Gans' formula for the dipole approximation\cite{Gans12}.
For the sample in Figure~\ref{fig:Image}, the full curve fit\cite{Slyusarenko:2013} of the absorbance by~\eqref{Eq:Abs} is very good (see Figure~\ref{fig:Abs}) and yields an order parameter $S = -0.15 \pm 0.01$.

\begin{figure*}[htbp]
\includegraphics[width=0.9\textwidth,angle=0]{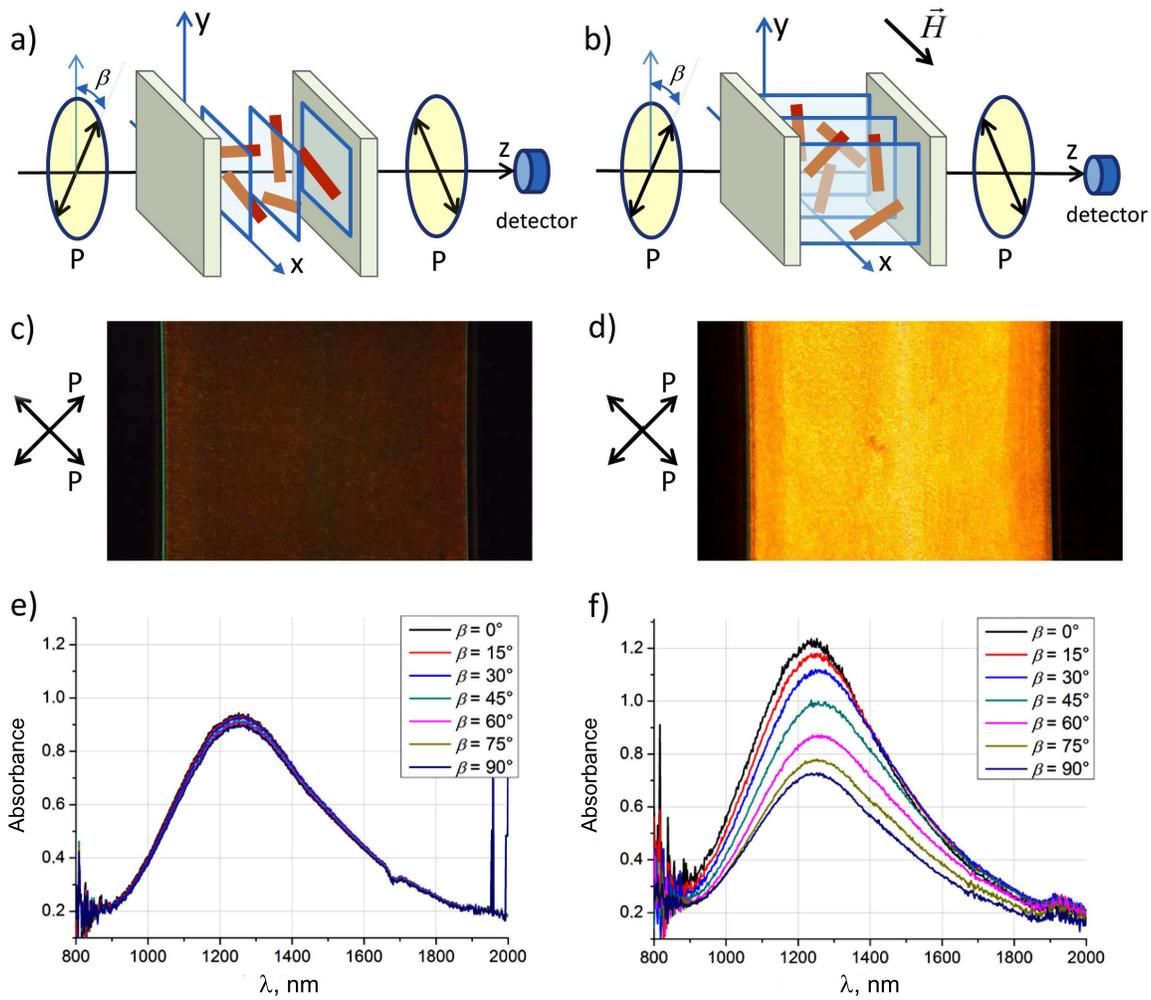}
\caption{Diagram of the experimental setup, with the lamellar phase in homeotropic (a) and planar orientation (b). The double arrows labelled ``P'' represent the directions of the crossed polarizer and analyzer and the angle between polarizer and the vertical axis $z$ is denoted by $\beta$. The magnetic field $\vect{H}$ is shown in (b). Polarized optical microscopy images (in the visible range) for situations (a) and (b) are shown in (c) and (d), respectively. The capillary width is 1~mm. Absorbance spectra of the logitudinal plasmon resonance peak in the infrared range for situations (a) and (b) are shown in (e) and (f), respectively. These measurements were made using a dedicated infrared polarizer (see Materials) and no analyzer. The optical path length is $50 \un{\mu m}$.}\label{fig:POM}
\end{figure*}

\subsection{Absorbance dichroism}

In order to exploit the anisotropy of the lamellar phase to orient the gold nanorods in the desired direction, its texture must be switched to planar orientation ($\vect{n}$ parallel to the faces of the capillary, see Figure~\ref{fig:POM}b). To this end, we further add to the lamellar phase goethite nanorods that confer magnetic sensitivity to the system, at a concentration $\phi_g = 1.5 \un{vol \%}$ (Note that $\phi_g$ is much higher than $\phi_\text{Au}$!). The goethite nanorods were thoroughly characterized in previous work \cite{Lemaire02,Lemaire04b}, and the lamellar phase doped with these nanoparticles (but without gold nanorods) was also investigated \cite{Beneut08,Constantin10}. . Hence, by doping the gold nanorod mesophase with magnetic nanoparticles, our aim is to couple the orientation of the two populations of nanorods through the surfactant mesophase and provide magnetic addressability to the gold nanorods.

This coupling occurs because both the goethite nanorods \cite{Slyusarenko:2013} and the gold nanorods (see Figure~\ref{fig:Abs}b and the discussion below) are confined within the water layers of the lamellar phase. Furthermore, it was shown that the orientation of the lamellar phase could be controlled by the application of a magnetic field, since the magnetic rods impose their orientation to the surfactant mesophase\cite{Beneut08}. Consequently, the magnetic field indirectly controls the orientation of the gold nanorods.

We measured the birefringence and dichroism of the sample as a function of the strength of the magnetic field, applied perpendicular to the light beam that was parallel to the sample normal, Figure~\ref{fig:POM}a. In the absence of the field, the lack of sample birefringence (Figure~\ref{fig:POM}c) indicates that the lamellar phase orients homeotropically and that the goethite nanorods have a uniaxial nematic distribution along the normal to the sample \cite{Beneut08,Constantin10}. For magnetic field intensities below $350~\un{mT}$, when the lamellar phase remains homeotropic, the dichroism was zero (Figure~\ref{fig:POM}e), \textit{i.e.} the director of the gold nanorods remains parallel to the beam.

Under stronger fields, above $400~\un{mT}$, the goethite nanorods reorient perpendicular to $\vect{H}$. The lamellar phase follows \cite{Beneut08}, so that $\vect{n}$ becomes parallel to $\vect{H}$ (and perpendicular to the beam), and the distribution of the gold nanorods tilts by $\pi/2$ with respect to the initial orientation, see Figure~\ref{fig:POM}a. Indeed, the sample showed strong birefringence (Figure~\ref{fig:POM}d) and negative dichroism $A_{\|} - A_{\bot} < 0 $, where $A_{\|}$ and $A_{\bot}$ are the absorbance values (measured at \lpp) corresponding to light polarization parallel and perpendicular to $\vect{n}$, respectively (Figure~\ref{fig:POM}f). We checked that the magnetic field has no measurable effect on the orientation of the gold nanorods by themselves, in water or in the lamellar phase.

This behavior indicates that, in the lamellar phase, the gold nanorods have a nematic uniaxial distribution with their director along the lamellar phase director. The order parameter can be obtained as (see \S~\ref{subsec:absorb}):
\begin{equation}
\label{Eq:Sexp}
S_{\text{exp}} = \cfrac{A_{\|} - A_{\bot}}{A_{\|} + 2 A_{\bot}} \, .
\end{equation}
From the experimental data in Figure~\ref{fig:POM}f we obtained  $S=-0.17 \pm 0.01$. This is remarkably close to the value of $-0.15 \pm 0.01$ estimated above for the gold nanorods under confinement in the lamellar phase, in homeotropic alignment and without goethite. We conclude that the presence of goethite does not perturb the distribution of the gold nanorods and that the magnetic field induces a complete orientation transition (from homeotropic to planar).

The absorbance spectra in Figure~\ref{fig:POM}e) and f) show a change in the position and shape of the longitudinal peak of the gold nanorods with respect to their spectrum in water (Fig.~\ref{fig:Abs}b) We attribute this to changes in dielectric properties of the medium due to the presence of the goethite nanorods but we have not studied this feature quantitatively. Note that the position of the longitudinal peak is exactly the same in water and in the lamellar phase (without goethite), see Figure~\ref{fig:Abs}b.

The goethite absorbs very strongly at the position of the transverse peak of the gold nanorods (around 520~nm), so that we cannot measure adequately the transverse peak of the gold nanorods in the presence of goethite. For this reason, in Figure~\ref{fig:POM} we only show the infrared region of the spectrum, where the goethite absorbance is negligible.

\section{Conclusions}

We have successfully aligned gold nanorods on macroscopic scales by confining them between the bilayers of a surfactant lamellar phase. Their negative order parameter was measured by fitting the absorbance spectrum.
By doping the phase with goethite nanorods, we can tune the orientation of the entire sample using a moderate magnetic field. Both the surfactant mesophase and the gold nanorods are dragged by the magnetic nanoparticles. This results in strong dichroism in the infrared range, which can be potentially tuned by varying the aspect ratio of the nanorods \textit{via} the synthesis conditions. This property can be of great interest for fabricating new optical elements addressable by an external field. 

\section{Acknowledgements}
We acknowledge support from the Triangle de la Physique (project 2011-083T).
\balance

\footnotesize{
\bibliography{Slyusarenko} 
\bibliographystyle{rsc} 
}

\end{document}